\documentclass[%
 reprint,
 superscriptaddress,
showpacs,preprintnumbers,
nofootinbib,
 amsmath,amssymb,
 aps,
prd,
]{revtex4-1}

\usepackage{graphicx}% Include figure files
\usepackage{dcolumn}% Align table columns on decimal point
\usepackage{bm}% bold math
\usepackage{hyperref}% add hypertext capabilities
\newcommand{\eqn}[1]{eq.~(\ref{#1})}

\newcommand{\eqns}[2]{eqs.~(\ref{#1})-(\ref{#2})}

 \newcommand{\beq}{\begin{equation}}
 \newcommand{\eeq}{\end{equation}}
 \newcommand{\beqa}{\begin{eqnarray}}
 \newcommand{\eeqa}{\end{eqnarray}}
 
 \def\lsim{\ \rlap{\raise 3pt \hbox{$<$}}{\lower 3pt \hbox{$\sim$}}\ }
 \def\gsim{\ \rlap{\raise 3pt \hbox{$>$}}{\lower 3pt \hbox{$\sim$}}\ }

 \def\chiu{{\raise 2pt \hbox{$\,\chi$}}{\lower 1pt \hbox{$_{u\,}$}}}
 \def\chiuqt{{\raise 2pt \hbox{$\,\chi$}}
            \hbox{$^4\!\!$}{\lower 1pt \hbox{$_{\!u\,}$}}}
\def\chiusq{{\raise 2pt \hbox{$\,\chi^2\!\!$}}{\lower 1pt \hbox{$_{u\,}$}}} 
\def\chid{{\raise 2pt \hbox{$\,\chi$}}{\lower 1pt \hbox{$_{d\,}$}}}
 \def\chidsq{{\raise 2pt \hbox{$\,\chi^2\!\!$}}{\lower 1pt \hbox{$_{d\,}$}}}
 \def\chie{{\raise 2pt \hbox{$\,\chi$}}{\lower 1pt \hbox{$_{e\,}$}}}
 \def\chiesq{{\raise 2pt \hbox{$\,\chi^2\!\!$}}{\lower 1pt \hbox{$_{e\,}$}}}
 \def\chiud{{\raise 2pt \hbox{$\,\chi$}}{\lower 1pt \hbox{$_{u,d\,}$}}}
 \def\chiude{{\raise 2pt \hbox{$\,\chi$}}{\lower 1pt \hbox{$_{u,d,e\,}$}}}

\begin{document}

\preprint{FTUAM-11-45 \quad   IFT-UAM/CSIC-11-26}

 \title{Naturally large Yukawa hierarchies }

 \author{Enrico Nardi}
\email{enrico.nardi@lnf.infn.it}
\affiliation{
INFN, Laboratori Nazionali di Frascati, %  \\ % [-6pt]
                Via Enrico Fermi 40,    I-00044 Frascati, Italy, \\  %% [2pt]
%} \affiliation{
and Departamento de F\'{\i}sica Te\'orica,  
                C-XI, Facultad de Ciencias,  \\ %  [-6pt]       
                Universidad Aut\'onoma de Madrid,       
                C.U. Cantoblanco, 28049 Madrid, Spain \\  %% [2pt]
%}\affiliation{
and Instituto de F\'{\i}sica Te\'orica UAM/CSIC,  \\ %  [-6pt]
Nicolas Cabrera 15, C.U. Cantoblanco, 28049 Madrid, Spain}

\begin{abstract}
The spontaneous breaking of the $SU(3)^5$ quark/lepton flavor symmetry
by means of three multiplets of scalar `Yukawa fields' admits vacua
with one ${\cal O}(1)$ and two vanishing vacuum expectation values
(vevs) for each multiplet.  If the number of generations is equal to
three, and only in this case, the vanishing vevs are lifted to
exponentially suppressed entries by the inclusion of symmetry
invariant logarithmic terms. A strong hierarchy for the Yukawa
couplings and a quark mixing matrix that approaches a diagonal form 
are obtained in a natural way from ${\cal O}(1)$ parameters.  This
scenario provides a concrete realization of the minimal flavor
violation hypothesis.
\end{abstract} 


%%%%%%%%%%%%%%%%%%%%%%%%%%%%%%%%%%%%%%%%%%%%%%%%%%%%%%%%%%%%%%%%%%%
   \maketitle
%%%%%%%%%%%%%%%%%%%%%%%%%%%%%%%%%%%%%%%%%%%%%%%%%%%%%%%%%%%%%%%%%%%

% \setcounter{page}{1} 
% \baselineskip 16pt 

\section{Introduction}

The Standard Model (SM) provides and accurate description of particle
physics phenomena. Particles interactions are derived from local
symmetries and are explained at a fundamental level by the gauge
principle. Myriads of experimental tests have confirmed the
correctness of this picture.  However, the SM cannot explain the
values of the particle masses and mixing angles. Even if the seed of
fermion masses is eventually identified, as it might happen soon at
the LHC, a new theory is required for explaining the puzzling features
of the observed pattern of Yukawa couplings: 1.~For each value of the
electric charge $Q=-1,\,-\frac{1}{3},\,+\frac{2}{3}$ there is a
threefold replica of fermions that are characterized by the same set
of (known) quantum numbers, and as a result the Yukawa couplings are
arranged into generic $3\times 3$ matrices. We have no clue about the
origin of fermion family replication.  2.~Unless new quantum charges
are postulated, fermions belonging to different generations but with
the same quantum numbers under the SM gauge group are
indistinguishable at the fundamental level.  It is then puzzling that
their masses are instead arranged in a strong hierarchical structure.
3.~The Yukawa matrices for the up and down quarks constitute two
sets of mutually independent parameters.  It is then surprising that
in the basis in which the down quark Yukawa matrix is diagonal and
with a given ordering of its entries e.g. from small to large,
the Yukawa matrix for the up quarks, when ordered in the same
way, is also approximately diagonal.  4.~Additionally, the theoretical
prejudice that there is new physics not too far above the electroweak
scale brings in one more puzzle: why new physics effects are not seen
in flavor violating processes?  We believe that the scenario we are
going to discuss can shed some light on all these issues.

\section{Symmetry and  renormalizable invariants}

Looking at the SM gauge sector one can readily recognize that fermions
are arranged into triplets of states with the same gauge quantum
numbers. It is then natural to postulate some symmetry group that
commutes with the SM gauge group and has three-dimensional
representations.  The symmetry, however, is not realized in the
spectrum, and generally this signals  a non invariant ground state
yielding spontaneous symmetry breaking (SSB). To pursue further these
simple considerations, we need to identify the symmetry group, and
make an ansatz about the way it is broken. A brief review of known
properties of the SM can guide us in carrying out this task.

The group of symmetry transformations of the SM quarks and leptons
gauge invariant kinetic terms is~\cite{MFV} ${\cal G}= {\cal
  G}^{(q)}\times {\cal G}^{(l)}$ with ${\cal G}^{(q)}=U(3)_Q\times
U(3)_u\times U(3)_d$ and ${\cal G}^{(l)}= U(3)_\ell\times U(3)_e$,
where $Q$ and $\ell$ denote the quark and lepton $SU(2)$ doublets, and
$u,\,d,$ and $e$ the quark and lepton $SU(2)$ singlets.  In the SM
${\cal G}$ is broken explicitly by the fermions Yukawa
couplings, however, some $U(1)$ factors are left unbroken.  In the
quark sector $U(1)_Y\times U(1)_B$ of hypercharge and baryon number
remain good symmetries, and in the lepton sector $U(1)_Y$ remains
unbroken as well.  The SM lepton sector is however incomplete since it
cannot accommodate massive neutrinos, and it is likely, although
experimentally not yet confirmed, that unlike baryon number $U(1)_L$
of lepton number is broken.  Whether this is true (Majorana neutrinos)
or not (Dirac neutrinos) is not relevant for our discussion, thus in
the following we assume that the broken subgroup of ${\cal G}$ is
${\cal G}_{\cal B}= {\cal G}_{\cal B}^{(q)}\times {\cal G}_{\cal
  B}^{(l)}$ with
\begin{eqnarray}
  \label{eq:GBq}
{\cal G}_{\cal B}^{(q)}&=&SU(3)_Q\times SU(3)_u
\times SU(3)_d\times U(1)_d\,,   \\
  \label{eq:GBl}
{\cal  G}_{\cal B}^{(l)} &=&SU(3)_\ell\times SU(3)_e\times U(1)_e\,. 
\end{eqnarray}
The Abelian factors $U(1)_{d,e}$ correspond to phase rotation of the
$SU(2)$ singlets $d$-quarks and $e$-leptons.  For the quarks, the
Abelian factor could have equally well chosen to be $U(1)_{u}$;
however, assuming that the Yukawa couplings of the $d$-quarks (and of
the leptons~\cite{Joshipura:2009gi,Alonso:2011jd}) break an additional
symmetry with respect to the Yukawa couplings of the $u$-quarks can
provide a simple justification for the suppression for the bottom (and
tau) mass with respect to the mass of the top. We will exploit this
symmetry argument in what follows.

Guided by the previous considerations, we assume a fundamental
symmetry that contains ${\cal G}_{\cal B}={\cal G}_{\cal
  B}^{(q)}\times {\cal G}_{\cal B}^{(l)}$ as a subgroup that gets
spontaneously broken, and that under the quark~\eqn{eq:GBq} and
lepton~\eqn{eq:GBl} factors the SM fermions transform respectively as
\begin{eqnarray}
  \label{eq:quarks}
Q &=&(3,\,1,\,1)_0\,,  \quad  
u\; =\;(1,\,3,\,1)_0\,,  \quad  
d\; =\;(1,\,1,\,3)_1\,,  \quad \\ 
  \label{eq:leptons}
\ell &=&(3\,,\,1)_0\,,  \quad \quad  
e \;=\;(\,1,\, 3)_1\,.  
\end{eqnarray}
As regards the way ${\cal G}_{\cal B}$ is dynamically broken, the
simplest choice is to interpret the SM explicit breaking as the result
of SSB.  That is, we assume that the Yukawa couplings of the SM quarks
and leptons correspond to vacuum expectation values (vev) of scalar
fields $Y_u$, $Y_d$, $N_d$ and $Y_e$, $N_e$ that are coupled to
the fermions in a symmetry invariant way via non-renormalizable
operators:
\begin{equation}
  \label{eq:nonren}
  -{\cal L}_Y = \frac{1}{\Lambda}\, \bar Q\, Y_u\, u\, H + 
\frac{1}{\Lambda^2}\, N_d\, \bar Q\, Y_d\, d\, \tilde H +
\frac{1}{\Lambda^2}\, N_e\, \bar \ell\, Y_e\, e\, \tilde H\,, 
\end{equation}
where $H$ is the Higgs field (with $\tilde H =i\sigma_2 H^*$) and
$\Lambda$ is a large scale where the effective operators in
\eqn{eq:nonren} arise. Invariance of ${\cal L}_Y$ under 
 ${\cal G}_{\cal B}^{(q)}$ and $ {\cal G}_{\cal B}^{(l)}$ fixes the 
following assignments:
\begin{eqnarray}
  \label{eq:scalarsq}
Y_u\! &=&\! (3\,,\bar 3,\,1)_0,  \   
Y_d = (3\,,\,1,\,\bar 3)_0,   \   
N_d = (1\,,1,\,1)_{-1},  \\
  \label{eq:scalarsl}
Y_e\! &=&\! (3\,,\bar 3)_0,  \    \quad
 N_e =(1\,,1)_{-1}.  
\end{eqnarray}
A more economical choice than the three multiplets of scalars in
\eqn{eq:scalarsq} and the two in \eqn{eq:scalarsl} is also
possible. In fact by assigning $U(1)_{d,e}$ charges to $Y_{d,e}$ we
would not need to introduce the complex scalars $N_{d,e}$.  It is,
however, more convenient to keep a clear distinction between the
hierarchy between the top and bottom/tau masses from the Yukawa
hierarchy between quarks and leptons of the same type.  For
simplicity, we will describe the first one by means of two
`Abelian spurions' $\eta_{N_{d,e}}\equiv \langle N_{d,e}\rangle/\Lambda$
and we take $\eta_{N_{d,e}} \approx m_{b,\tau}/m_t $ as given numbers. We
will briefly comment on the SSB of $U(1)_{d,e}$ only at the end of the
paper. The hierarchy between generations is instead explained via SSB of
the $SU(3)^5$ quark/lepton flavor symmetry, that is by the dynamical
selection of vevs with the required structure.  

We note, in passing, that the introduction of the Abelian spurion
$\eta_{N_d}$ implies that $U(1)_d$ invariant operators like $\bar Q\,
Y_dY_d^\dagger\, Q$ are not suppressed with respect to $\bar Q\,
Y_uY_u^\dagger\, Q$.  This is different from what is commonly assumed
in Minimal Flavor Violation (MFV) extensions of the
SM~\cite{DAmbrosio:2002ex}, in which effective operators involving
$Y_d$ are always suppressed, and it resembles more MFV extensions of
two Higgs doublets models like supersymmetry~\cite{DAmbrosio:2002ex}
in which $\tan\beta$ plays basically the role of $\eta_{N_{d,e}}$. The
formal difference is that in the present case the absence of
suppression factors for $Y_dY_d^\dagger$ follows from a symmetry
argument.

The assignments in~\eqns{eq:quarks}{eq:leptons} imply that all the
fermions with the same quantum numbers under the SM gauge group are
characterized by the same quantum numbers also under ${\cal G}_{\cal
  B}$, and therefore the different generations contain exact replica
of the same set of states.  As we will see, the fact that for each
triplet of identical fermions two Yukawa couplings in first
approximation vanish while the third one is ${\cal O}(1)$, corresponds
precisely to a non symmetric ground state that yields a symmetry
breaking pattern in qualitative agreement with observations.  Note
that this picture is fundamentally different from assuming new
symmetries under which fermions with the same SM quantum numbers
transform differently, that is for example the basic ingredient of the
popular Froggatt-Nielsen mechanism~\cite{Froggatt:1978nt}.  In that
case the hierarchy of the Yukawa couplings follows from a dimensional
hierarchy in the corresponding effective Yukawa operators, that is
obtained by assigning to the lighter generations larger values of new
Abelian charges. The fact that fermion families appear to replicate
would then be just an illusory feature due to our incomplete knowledge
of the fundamental symmetries, and not a fundamental property of
the SM fermions.

\subsection{Symmetry invariants and the TAD parametrization}
To carry out a basis independent analysis of the SSB of the $SU(3)^5$
flavor symmetry, it is convenient to write the multiplets of scalar
`Yukawa fields' $Y_{u,d,e}$ in their singular value  decomposition:
\begin{equation}
  \label{eq:biunitary}
  Y_u = {\cal V}_u^\dagger\, \chiu\,  {\cal U}_u \,,\  \ 
  Y_d =  {\cal V}_d^\dagger\, \chid\,  {\cal U}_d \,,\  \ 
  Y_e =  {\cal V}_e^\dagger\, \chie\,  {\cal U}_e \,,
\end{equation}
where the matrices $ {\cal V}$ and $ {\cal U}$
are unitary  while  the matrices $\chi$ are diagonal with nonnegative real
entries:
\begin{eqnarray}
\nonumber
  \label{eq:diag}
\chiu &=& {\rm diag}\left(u_1,\,u_2,\,u_3\right)\,, \quad  
\chid \  =\  {\rm diag}\left(d_1,\,d_2,\,d_3\right)\,,\\   
\nonumber
\chie  &=& {\rm diag}\left(e_1,\,e_2,\,e_3\right)\,.   
\end{eqnarray}
Note that while the matrices of {\it singular values} $\chi$ are
unique (modulo reordering of their entries) $ {\cal V}$ and $ {\cal
  U}$ are not, and can be redefined according to $ {\cal V}\to \Phi
{\cal V} $, $ {\cal U}\to \Phi {\cal U} $ where $\Phi$ is a diagonal
matrix of phases. This can be used to remove 3 phases for example
in $ {\cal V}$.

We now  require invariance of the scalar potential for the Yukawa fields 
under the special bi-unitary transformations $Y_{u,d} \to V_Q\,Y_{u,d}\,
U_{u,d}^\dagger$ and $Y_{e} \to V_\ell\, Y_{e}\, U_{e}^\dagger$.
Assuming three generations, for the $u$-quark sector we can
write the following invariants:
\begin{eqnarray}
\label{eq:Tu}
  T_u &=& {\rm Tr}( Y_uY_u^\dagger)= \sum_i u_i^2\,, \\
\label{eq:Au}
  A_u &=& {\rm Tr}\left[{\rm Adj}( Y_uY_u^\dagger)\right]= 
\sum_{i\neq j} u^2_iu^2_j\,,\\  
\label{eq:Du}
  {\cal D}_u &=& {\rm Det}(Y_u)= e^{i\delta_u}\,\prod_i u_i \equiv e^{i\delta_u}\,D\,, 
\end{eqnarray}
where $\delta_u= {\rm Arg}\left[{\rm Det}\left({\cal V}_u^\dagger{\cal
      U}_u\right)\right]$, and a similar relation holds for ${\cal
  D}^*_u = {\rm Det}(Y_u^\dagger)$ with $\delta_u \to -\delta_u$.
The invariance of the trace $T_u$ and of the determinant ${\cal
  D}_u$ under a $SU(3)_Q\times SU(3)_u$
transformation $Y_u \to V_Q\, Y_u \, U_u^\dagger$ with the special
unitary matrices ${\rm Det}(V_Q)={\rm Det}(U_u^\dagger)=+1$ is
obvious.  The second invariant in~\eqn{eq:Au} is the trace of the {\it
  adjugate}, that is the trace of the transpose of the matrix of
cofactors.  Its invariance can be proved as follows: Laplace's formula
applied to the product of two $n\times n$ matrices $A$ and $B$ reads
$(AB) \cdot {\rm Adj}(AB) = {\rm det}(AB)= {\rm det}(A)\cdot {\rm
  det}(B)$ from which the product rule ${\rm Adj}(AB)= {\rm
  Adj}(B)\cdot {\rm Adj}(A)$ is easily derived. Moreover, for a
unitary matrix ${\rm Adj}(V)=V^\dagger$. Invariance of $A_u$ under $
Y_uY_u^\dagger \to V_Q\, Y_uY_u^\dagger\, V_Q^\dagger$ then follows
straightforwardly. For general $n\times n$ matrices $T_{uu} = {\rm
  Tr}( Y_uY_u^\dagger Y_uY_u^\dagger )= \sum_i u_i^4$ is also a
renormalizable invariant.  However, in the case of $3\times 3$
matrices it is not an independent one: $T_{uu} =T_u^2 - A_{u}$.  Other
two sets of $T,A,{\cal D}$ invariants completely similar
to~\eqns{eq:Tu}{eq:Du} can be written also for $Y_d$ and $Y_e$, and
thus all the results that we will derive for $Y_u$ apply equally well
also to the them.  Therefore, in the following we will drop wherever
possible the subscript $u$.

\section{Symmetry Breaking}
In this section we study the general potential for the Yukawa fields
$Y_{u,d,e}$ invariant under ${\cal G}_{\cal B}$, and we classify the
different minima that yield SSB.  Some of these issues have been
recently addressed also  in~\cite{Feldmann:2009dc,Alonso:2011yg}. The
renormalizable potential constructed from the $T,A,{\cal D}$
invariants~\eqns{eq:Tu}{eq:Du} reads:\footnote{According to our
  simplification of betraying the scalars $N_{d,e}$ for the two
  spurions $\eta_{N_{d,e}}$ we initially omit writing $NN^\dagger T$
  terms. A coupling with the Higgs $H H^\dagger T$ can also be omitted
  as long as $\langle H H^\dagger \rangle/\Lambda^2 \ll m^2$.}
\begin{eqnarray}
  \label{eq:V3}
  \hat V &=&  \Lambda^4\,V 
= \Lambda^4\left( V_{T} + V_{A}+V_{\cal D}\right) \,,  \\   
  \label{eq:VT3}
V_{T} &=&\lambda \left[T- \frac{m^2}{2\lambda}\right]^2\!, \\
  \label{eq:VA3}
  V_A &=&  \tilde\lambda' A\,, \\ 
  \label{eq:VD3}
V_{\cal D} &=& \tilde \mu\, {\cal D} +  \tilde \mu^*\, {\cal D}^* 
=2\,\mu\,\cos(\phi+\delta)\,D \,. 
\end{eqnarray}
In the first equation we have factored out a large scale $\Lambda$,
that for simplicity can be identified with the effective scale
in~\eqn{eq:nonren}, so that all the parameters and fields in
$V=V_T+V_A+V_D$ have no dimensions, and for definiteness we also
assume that all the Lagrangian parameters are renormalized at this
same scale.  $V_T$ in~\eqn{eq:VT3} contains the two invariants
constructed from the trace $V_T=\lambda T^2 -m^2 T$ plus an irrelevant
constant.  We require $\lambda>0$ and $m^2 > 0$ in order to have a
potential bounded from below and to trigger SSB.  The last equality
in~\eqn{eq:VD3} is obtained by defining $\tilde \mu = \mu e^{i\phi}$
that is $\mu\equiv |\tilde \mu|=|\tilde \mu^*|$.  The parameter
$\tilde\lambda'$ that multiplies $A$ can be either positive or
negative, and we need to consider both possibilities.  In the
following we will refer to its absolute value
by simply dropping the tilde $\lambda'\equiv
|\tilde\lambda'|$.

By exploiting the TAD parametrization introduced
in~\eqns{eq:Tu}{eq:Du} the hierarchy of the SM Yukawa couplings can be
conveniently described in terms of vevs of invariants, and corresponds
to minima that satisfy the condition:
\begin{equation}
   \label{eq:qualitative}
 \langle D \rangle ^{\frac{1}{3}} \ll 
\langle A\rangle^{\frac{1}{4}}  \ll \langle T\rangle^{\frac{1}{2}}\,.  
\end{equation}
Of course, in our dimensionless approach the exponents can be equally
well dropped.  There is a simple correspondence between the TAD vevs,
and the vevs of the field components $\langle u_i\rangle$.  For
example, assuming that a vev $\langle \chi\rangle $ with a large
hierarchy for its components has been found, by labeling its entries
according to $\langle u_1\rangle \ll \langle u_2\rangle \ll \langle
u_3\rangle$ we have:
\begin{equation}
  \label{eq:hierarchy}
  \frac{\langle A\rangle}{\langle T\rangle^2 }
\approx \frac{\langle u_2^2\rangle }{\langle u_3^2\rangle}\,,\qquad \qquad 
  \frac{\langle D\rangle^2}{\langle T\rangle \langle A\rangle }
\approx \frac{\langle u_1^2\rangle }{\langle u_3^2\rangle}.
% \frac{\langle T\rangle\,\langle D\rangle^2}{\langle A\rangle^2 }
% \approx \frac{\langle u_1^2\rangle }{\langle u_2^2\rangle}.
\end{equation}
Since the value of the top Yukawa coupling fixes $\langle T_u\rangle
\approx \langle u_3^2\rangle \approx 1$ and naturalness suggests   
$\langle T_{d,e}\rangle\approx {\cal O}(1)$ as well, 
% \eqn{eq:qualitative} suggests that 
it follows that vacua characterized in first approximation by $\langle
D \rangle= \langle A \rangle =0 $ and $\langle T\rangle \approx 1$ are
well suited for generating the Yukawa hierarchies.

From \eqn{eq:VT3} we immediately see that $V_T$ is minimized for
values on the spherical surface in $\langle u_i\rangle$ space $\langle
T\rangle= \frac{m^2}{2\lambda}$.  Note that while we must require
$\frac{m^2}{2\lambda}\approx 1$, a perturbative $\lambda<1$ implies
$m^2 < 1$, and thus in first approximation contributions of non
renormalizable operators to $V$ can be neglected.  As regards $A$ and
$D$, they are both maximized for symmetric vacua $\langle \chi \rangle
=(u_s,\,u_s,\,u_s) $ and minimized when they vanish.  To ensure
$\langle {\cal D}\rangle=0$ either 
$\langle \delta\rangle=\pm \frac{\pi}{2}-\phi$ or
at least one entry in $\chi$ must vanish, while for $\langle A\rangle
=0$ two entries must vanish.  Which particular minimum on the surface
$\langle T \rangle=\;$const is selected then depends on the signs and
values of $\tilde\lambda'$ and of $\mu\cdot\cos(\phi+\delta)$.  Below we
classify the different types of minima yielding SSB, and we show that
vacua with the required property $\langle D\rangle =\langle A\rangle
=0$ indeed occur.

\smallskip 

{\it Case 1:\ $\tilde\lambda'=-\lambda'<0$.}\quad
$V_A$ is negative and its absolute value is maximized for symmetric
vacua $\langle \chi\rangle =(u_s,\,u_s,\,u_s)$. A negative value of
$V_{\cal D}$ lowers further the minimum, which fixes $\langle
\delta\rangle =\pi-\phi$, while $D$ is also maximized for symmetric
vacua. The potential is bounded if $\lambda'<3\lambda$, in which case
we obtain:
\begin{equation}
  \label{eq:Case1}
  u_s = \frac{\mu}{4(3\lambda-\lambda')}\left[1+\sqrt{1+8
(3\lambda-\lambda')\,\frac{m^2}{\mu^2}}\;\right]\!,  
\end{equation}
This case yields a phenomenologically uninteresting non-hierarchical
pattern $\langle T\rangle\approx\langle D\rangle\approx \langle
A\rangle\approx 1$.

\smallskip

{\it Case 2:\  $\tilde\lambda'=\lambda'>0$.}\quad
In this case $V_A=\lambda A$ is positive and is minimized when $A=0$,
which favors vacua with two vanishing entries $\langle \chi \rangle =
(0,0,u_t)$.  The value of $V_{\cal D}$ is extremized either when:
\\  \indent 
{\it Case 2a:} $V_{\cal D}=-2\mu D$, which occurs
for $\langle \delta\rangle=\pi-\phi$. In this case $\langle \chi
\rangle$ acquires a symmetric structure $(u_s,u_s,u_s)$ that
maximizes $D$. Or when: \\ \indent 
{\it Case 2b:} $\langle \chi
\rangle$ has at least one vanishing entry yielding $V_{\cal D}=0$. In
this case $\langle \delta\rangle$ is left undetermined.  $\langle
\chi \rangle = (0,0,u_t)$ is in fact favored over a single vanishing entry
because it also ensures $V_A=0$. The two vevs are:
\begin{eqnarray}
  \label{eq:case2a}
  u_s &=& \frac{\mu}{4(3\lambda+\lambda')}\left[1+\sqrt{1+8
(3\lambda+\lambda')\,\frac{m^2}{\mu^2}}\;\right]\!, \\ 
  \label{eq:case2b}
u_t &=& \frac{m}{ \sqrt{2\lambda}}\,. 
\end{eqnarray}
Since  $V(u_t)=0$,  vacua with two vanishing entries are selected  
if $V(u_{s})$ is positive, which occurs for 
\begin{equation}
  \label{eq:sol3}
  \frac{\mu^2}{m^2}<2 \lambda
\left[\left(4+\frac{\lambda'}{\lambda}\right)^{\frac{3}{2}}
-\left(8+3\frac{\lambda'}{\lambda}\right)\right]\,. 
\end{equation}
In this case $\langle T\rangle\approx 1$ while $\langle D\rangle
=\langle A\rangle =0$ which represents an interesting first
approximation to~\eqn{eq:qualitative}.

\subsection{Lifting the  vanishing vevs} 

In Cases 2b the vacua are characterized by $\langle D \rangle=\langle
A \rangle=0$ which is a good starting condition to generate minima
satisfying~\eqn{eq:qualitative}.  Of course, to have an acceptable
phenomenology we must lift these two vevs to appropriately small, but
non-vanishing values. In order to achieve this let us add to $V$ an
invariant term proportional to the logarithm of the three-point
interaction:
\begin{equation}
\label{eq:VLogD}
V_{\cal D}  \to  V_{\cal D} + V_{\cal D}^{(1)} =  
2 \mu \cos(\phi+\delta)\,D \left(1+c_D\log D\right)
% \tilde \mu {\cal D}\, (1+  c_{D} \, \log {\cal D}) + {\rm h.c.} 
% \qquad\qquad \qquad 
% \\  &&\hspace{-4mm} \nonumber
% = 2 \mu D\left[\left(1+c_D\log D\right)\,\cos(\phi+\delta)
% -\delta\,\sin(\phi+\delta)\right],\qquad    
\end{equation}
where $c_D$ is a small numerical factor.  Here, at the cost of some
arbitrariness, we justify the introduction of $V_{\cal D}^{(1)}$ only
on the basis of symmetry properties and of naive dimensional analysis;
however, we can expect that terms of this type will be generated by
quantum corrections to the effective potential, in which case the loop
coefficient $c_D$ would be computable.  The effect of $V_{\cal
  D}^{(1)}$ is that of generating a large negative log that shifts the
term $D(1+c_D \log D)$ towards a non vanishing negative value.  The
minimum is then obtained for $\langle \delta \rangle =-\phi$ and for a
non vanishing, but exponentially suppressed value of $D$:
\begin{equation}
  \label{eq:LogD}
 \langle D\rangle = e^{-\left(\frac{1}{c_D}+1\right)} \equiv \epsilon_D\,. 
\end{equation}
This gives an elegant realization of an old remark by
Nambu~\cite{Nambu:1992ke}.  Given that in the unperturbed solution
$\langle \chi \rangle =(0,0,u_t)$ we have $u_t\approx 1$,
\eqn{eq:LogD} implies $\langle u_1\rangle \cdot \langle u_2\rangle
\equiv u_u\cdot u_c \approx \epsilon_D$.  Note that  $\log D$ 
is unable to differentiate between $u_u$ and $u_c$, and that terms
involving higher powers of $\log D$ are also ineffective for
distinguishing these two entries.  However, because of the presence of
$V_A$ two different types of vacua are possible.  In case~4 $V_A>0$
and its smallest possible value $\langle A\rangle \sim \epsilon_D$ is
obtained when one component field has a vanishing vev: $\langle \chi
\rangle \sim (0,\epsilon_D,1)$.  In case 3 instead, $V_A<0$ favors
semi-symmetric vacua with $\langle \chi \rangle \sim
(\sqrt{\epsilon_D},\sqrt{\epsilon_D},1)$ corresponding to the largest
possible value $\langle A\rangle \sim 2\sqrt{\epsilon_D}$. Of course,
both these possibilities are phenomenologically untenable.  However,
adding a $\log A$ term makes the magic that in both these cases an
exponential hierarchy is induced also between the two suppressed
entries $u_u$ and $u_c$.
\begin{equation}
  \label{eq:VLogA}
V_A \to V_A + V_A^{(1)} =  \tilde\lambda'_A\, A (1+  c_{A} \, \log A)
\end{equation}
in fact yields:
\begin{equation}
  \label{eq:LogA}
 \langle  A \rangle = e^{-\left(\frac{1}{c_A}+1\right)} \equiv \epsilon_A\,.
\end{equation}
Then, if we assume $c_D<c_A$ the hierarchical pattern in
\eqn{eq:qualitative} is realized.  In terms of vevs of component
fields, setting for simplicity $u_t = 1$, eqs.~(\ref{eq:LogD})
and (\ref{eq:LogA}) give a system of two equations
\begin{equation}
  \label{eq:uuuc}
  u_u\cdot u_c = \epsilon_D\,,\qquad\qquad
 u_u^2 + u_c^2 = \epsilon_A- \epsilon_D^2\,,
\end{equation}
that has real solutions for $\epsilon_D\leq \epsilon_A/2$. At first
order in $\epsilon_D/\epsilon_A$ the solutions are:
\begin{equation}
   \label{eq:u1u2}
u_u^2 = \frac{\epsilon_D^2}{\epsilon_A}\,,\qquad\qquad
u_c^2 = \epsilon_A\,,  
\end{equation}
where the particular labeling  $u$ and $c$ is of course
arbitrary. Equivalently, inserting (\ref{eq:LogD}) and (\ref{eq:LogA})
in \eqn{eq:hierarchy} we obtain:
\begin{equation}
  \label{eq:u1u2approx}
  \frac{u_c^2}{u_t^2} \approx e^{-\left(\frac{1}{c_A}+1\right)}\,, 
  \qquad\qquad 
  \frac{u_u^2}{u_c^2} \approx e^{-2\left(\frac{1}{c_D}-
      \frac{1}{c_A}\right)}\,. 
\end{equation}
Without introducing any unnaturally small parameter we thus obtain a
pattern of Yukawa couplings that is characterized by an exponentially
strong hierarchy.  For the up-quark sector for example we need
$c_A\sim -\left(\log\langle A_u \rangle \right)^{-1}\approx0.10$ and 
$c_D\sim-\left(\log\langle D_u\rangle\right)^{-1}\approx
0.06$.\footnote{Because the number of field components in
  $Y$ is large: $N=18$, a typical loop factor
  $\frac{N}{64\pi^2}\approx 0.03$ is of about the correct size.}
Finally, we should also add to the $\lambda T^2$ interaction a 
$\log T$ correction.  This, however, has no particular consequences
since in any case the minimization of $V_T$ fixes $\langle T\rangle \approx 1$.

\subsection{Coupling the up and down quark sectors}

As long as the up and down quark sectors are treated separately, the
relative hierarchical ordering of the vevs of the component fields
$\langle u_i\rangle $ and $\langle d_i\rangle$ is irrelevant, and we
have a set of $3!\times 3!$ equivalent vacua. This large degeneracy is
partially reduced when the renormalizable invariants that
couple the two sectors are included. There are two possible terms:
\begin{eqnarray}
  \label{eq:TuTd}
T_u\cdot T_d&=& {\rm Tr}( Y_uY_u^\dagger)\cdot  {\rm Tr}(Y_dY_d^\dagger )
\ =\ \sum_{ij} u^2_i d^2_j \,, \\ 
  \label{eq:Tud}
T_{ud}  &=& {\rm Tr}( Y_uY_u^\dagger Y_dY_d^\dagger )\ =\ 
 {\rm Tr}\left(V^\dagger \chiusq  V  \chidsq \right) \,. 
\end{eqnarray}
where $V$ is a unitary matrix of fields that in terms of the
bi-unitary parametrization \eqn{eq:biunitary} is given by $V={\cal
  V}_u {\cal V}_d^\dagger$.  It is now convenient to replace $T_{ud}$ by
the invariant combination:
\begin{equation}
  \label{eq:Aud}
 A_{ud}\equiv T_u\cdot T_d- T_{ud} \,.
\end{equation}
Note that $A_{ud}$ is not related to adjugate matrices,
and its invariance follows solely from its definition.  The
contribution to the scalar potential from coupling the $u$ and $d$
sectors is
\begin{equation}
  \label{eq:Vprimeud}
  V_{ud} = 2 \tilde\lambda_{ud}\, T_u\cdot T_d \ +\ \tilde\lambda'_{ud} 
A_{ud}\ +\ \dots \,,
\end{equation}
where the dots stand for logarithmic terms, that we expect could be
relevant (see below) but whose effects in this case are difficult to
analyze analytically and thus, as a first approximation, we leave them
out.  We now show that the first term in $V_{ud}$ yields the same
vacuum conditions implied by~\eqn{eq:VT3}, that is that the minimum of
the potential is localized on the surface of two three spheres of
constant $\langle T_u\rangle$ and $\langle T_d\rangle$.  We start by
shifting the couplings of the terms linear and quadratic in $T_{u,d}$
(see \eqn{eq:VT3}) according to:
\begin{eqnarray}
  \label{eq:redefm}
m^2_{u,d} &\to&  m^2_{u,d} +  \tilde m^2_{ud} \,,\\ 
  \label{eq:redeflambda}
\lambda_{u,d}  &\to& \lambda_{u,d} + \tilde\lambda_{ud}\,, 
\end{eqnarray}
where $\tilde m^2_{ud}$ in the first line is an arbitrary constant of
the same sign than $\tilde\lambda_{ud}$ that we will fix in a moment.
With a little algebra, and omitting irrelevant constants, we can rewrite:
\begin{eqnarray}
  \label{eq:VTud}
V_{T_{(u+d)}}&\equiv&  
V_{T_u} + V_{T_d} + 2 \tilde\lambda_{ud}\, T_u\cdot T_d = 
% \lambda_u \left[T_u - \frac{m^2_{u}}{2 \lambda_{u}}\right]^2 
 \\
\nonumber
&&\hspace{-1.5cm}
 \lambda_u\! \left[T_u - \frac{m^2_{u}}{2 \lambda_{u}}\right]^2
\!\!\!+\! 
\lambda_d\! \left[T_d- \frac{m^2_{d}}{2 \lambda_{d}}\right]^2 
% \\  &&\hspace{-1.5cm}
\!\!\!+\!  
\tilde\lambda_{ud}\! 
\left[T_u+T_d- \frac{ m^2_{ud}}{2 \lambda_{ud}}\right]^2\!\!.
\end{eqnarray}
Note that the ratio $\tilde m^2_{ud}/\tilde \lambda_{ud}$ that
would appear in the last square bracket is always positive by
construction, which justifies omitting the tilde on both parameters.
We can now fix $m^2_{ud}$ to satisfy:
\begin{equation}
  \label{eq:mud}
 \frac{m^2_{ud}}{\lambda_{ud}}=  \frac{m^2_{u}}{\lambda_u}+
\frac{m^2_{d}}{\lambda_d}\,, 
\end{equation}
where $\lambda_{u,d}$ and $m^2_{u,d}$ are the redefined parameters
appearing in the r.h.s of~\eqns{eq:redefm}{eq:redeflambda}.  If
$\tilde\lambda_{ud}> - \frac{\lambda_u\cdot \lambda_d}{\lambda_u+
  \lambda_d}$\  SSB occurs, with the minimum of the potential located on
the two surfaces $\langle T_{u}\rangle= \frac{m^2_{u}}{2 \lambda_{u}}$
and $\langle T_{d}\rangle = \frac{m^2_{d}}{2 \lambda_{d}}$.  Note that
also the mixed $T_{u,d}\cdot T_e$  terms can be  `reabsorbed'
in a similar way.  Genuinely new effects, and in particular the relative
ordering of the Yukawa hierarchies of the $u$ and $d$ sectors, come
from the second term in~\eqn{eq:Vprimeud} that can be written more
explicitly as:\footnote{Using unitarity $\sum_j |V_{ij}|^2=\sum_j
  |V_{ji}|^2=1$ we can also write $ A_{ud}-\sum_{i\neq j}u^2_i d^2_j=
  \sum_{ij}u^2_i(d^2_i-d^2_j)\,|V_{ij}|^2=
  \sum_{ij}d^2_i(u^2_i-u^2_j)\,|V_{ji}|^2\,$ which puts in evidence
  that if $\langle d_i\rangle = \langle d_j\rangle$ or $\langle
  u_j\rangle = \langle u_i\rangle$ the corresponding $\langle V_{ij}
  \rangle$ remains undetermined.}
\begin{equation}
  \label{eq:Audnew}
\tilde \lambda'_{ud}  A_{ud}  =\tilde \lambda'_{ud}\,\sum_{ij}
  \left(1-|V_{ij}|^2\right) u^2_i d^2_j\,.
 \end{equation}
 Since $V$ is unitary, $|V_{ij}|^2\leq 1$, and $A_{ud}$ then is the
 sum of positive semi-definite terms that cannot all vanish, which
 ensures $A_{ud}>0$. For $\tilde\lambda'_{ud}=\lambda'_{ud}>0$ the
 contribution to the potential is then minimized when $A_{ud}$ is at
 its minimum and, as we will now show, this occurs when $\langle
 V_{ij}\rangle$ approaches a diagonal form.  Let us consider the
 vacuum configurations obtained in case 3 or 4 and label the two
 largest component vevs as $ u_{t} \approx d_{b} \sim {\cal
   O}(1)$. All the other vevs are exponentially suppressed and
 generically of ${\cal O}(\epsilon)$.  We can then write
\begin{equation}
  \label{eq:VAudnew}
  A_{ud}= \left(1-|V_{tb}|^2\right) 
\times   {\cal O}(1)
% \ \; 
  + 
\!\!\!
\sum_{(ij)\neq (tb)} 
\!\!\!
\left(1-|V_{ij}|^2\right)
\times   {\cal O}(\epsilon)\,. 
\end{equation} 
Small values $A_{ud} \sim {\cal O}(\epsilon)$ can result only when the
modulus of the field $V_{tb}$ that couples to the entries in $\chiu$
and $\chid$ that have the largest vevs is exponentially close to one,
and thus the minimum must occur around this configuration.  Unitarity
then implies that the off-diagonal vevs $\langle |V_{tj}|^2\rangle $
($j\neq b$) and $\langle |V_{jb}|^2\rangle $ ($j\neq t$) are
accordingly suppressed. In this way a `third generation structure'
emerges, in the sense that $t$ and $b$ get almost decoupled from the
other quarks.  For the two lighter generations it is more difficult to
carry out this argument: we would need to confront various different
contributions to $A_{ud}$ that are all of ${\cal O}(\epsilon)$, and
thus we are not allowed to neglect neither $V^{(1)}_{{\cal D},A}$ that
are also of ${\cal O}(\epsilon)$, nor a $\log T_{ud}$ term that can be
expected to induce corrections of a similar size. However, it is also
clear that at this level we do not have enough information to
determine univocally the structure of $\langle V\rangle$. For example
the fact that $V_{ij}$ enter \eqn{eq:VAudnew} only through their
modulus square implies that there is no information on complex phases.

To summarize, we have seen that once the component vevs in $\langle
\chiu\rangle $ and $\langle \chid\rangle $ are conventionally ordered
in the same way, e.g. with increasing size, minimizing the coupling
term $A_{ud}$ pushes $\langle V\rangle$ to approach a diagonal form, a
dynamical behavior that can provide an explanation for the most
puzzling feature of the CKM matrix.  Although only qualitative, this
result is certainly non-trivial, in fact a priory nothing could have 
guaranteed that the largest mixings do not occur between the ${\cal
  O}(1)$ and the ${\cal O}(\epsilon)$ components in $\langle
\chiud\rangle$, and actually something like this would occur for
$\tilde\lambda'=-\lambda'<0$.

\section{Why three generations ?}

Our approach to explain the fermion mass hierarchy does not provide an
explanation of why there are three families; however, it does imply an
interesting connection between the number of generations $n_g$ and the
qualitative features of the Yukawa couplings hierarchy.  Since this
section contains mainly simple arguments based on dimensional
analysis, we simplify the discussion by considering only the case of a
real determinant ${\cal D}=D$.

 For $n_g>4$, $A$ and $D$ have Dim$(A,D)>4$ and thus
cannot appear in the renormalizable potential.  However, $T_{uu}={\rm
  Tr}\chiuqt$ is now independent of $T_u$ and $A_u$ and provides a new
renormalizable invariant term. The potential can be conveniently
written as $V=\lambda\left(T-\eta^2\right)^2+\tilde\lambda'\sum_{i\neq
  j} u_i^2u_j^2$ where the first term yields $\langle T\rangle
=\eta^2$ while (for $\tilde\lambda'=\lambda'>0$) minimization of the
second term implies that $n_g-1$ components have vanishing vevs.  We
see that the absence of additional renormalizable interactions does
not allow neither lifting the vanishing determinant nor generating a
hierarchical pattern for the $n_g-1$ vanishing couplings.  

If $n_g=4$, then ${\rm Dim}(D)=4$ and we can arrange for an
exponentially suppressed $\langle D\rangle\neq 0$.  However, ${\rm
  Dim}(A)=6$ and then besides $T_u^2,\,T_{uu}$ and $D$ there are no
other renormalizable interactions. Therefore, also in this case it is
not possible to remove all the degeneracies between the suppressed
vevs and obtain a fully hierarchical pattern.

The case $n_g=2$ is a bit more involved and requires a more detailed
discussion.  For $2\times 2$ matrices $A_u=T_u$ and $T_{uu}=
T_u^2-2D_u^2$ so we have just the two basic invariants $T$ and $D$.
We can write the renormalizable potential as:
\begin{eqnarray}
  \label{eq:V2}
\hat V &=& \Lambda^4\,V 
= \Lambda^4\,\left(V_T+V_D\right)\,,\\ 
  \label{eq:VT2}
  V_T &=& \lambda\left(T -  \frac {m^2}{2\lambda}\right)^2\!,     
% \\ 
%  \label{eq:VD2}
\quad  V_D \ =\   
\tilde\lambda' \left(D +  \frac{\tilde \mu^2}{2\tilde\lambda'}\right)^2    
\!.
\end{eqnarray}
SSB occurs for $\lambda>0$ and $m^2>0$ and, if
$\tilde\lambda'=\lambda'>0$ and $\tilde\mu^2=\mu^2>0$ the determinant
vanishes at the minimum, implying a vanishing vev for one component
field. We can try to lift  $\langle D\rangle =0$  by adding to
the $\lambda' D^2$ interaction   a logarithm:
\begin{equation}
  \label{eq:V2Log}
V_D\to V_D+  V_D^{(1)} = \mu^2 D+\lambda' D^2(1+ c_D\, \log D) \,.  
\end{equation}
Varying  with respect to $D$ and equating to zero gives the condition 
\begin{equation}
  \label{eq:min2}
  \log \langle D\rangle = - \left(\frac{1}{c_D}+\frac{1}{2}\right)- 
\frac{\mu^2}{2\lambda'c_D}\,\frac{1}{\langle D\rangle}\equiv 
-\alpha-\frac{\beta}{\langle D\rangle}\,. 
\end{equation}
We have a minimum if $\beta \leq \langle D\rangle \ll 1 $ where the
second inequality, which implies $\frac{1}{c_D}\ll
\frac{2\lambda'}{\mu^2}\sim {\cal O}(1)$, follows from requiring a
large hierarchy.
% (presumably) both $\alpha,\,\beta \gg 1$. 
Let us  seek a solution of the form
\begin{equation}
  \label{eq:sol0}
\langle D\rangle = e^{-(\alpha-W)}
\end{equation}
with $W$ a suitable function.  Substituting this formal solution
into~\eqn{eq:min2} yields
\begin{equation}
  \label{eq:solW}
 W e^W = -\beta e^\alpha\,.
\end{equation}
The function $W$ defined implicitly through $W(x)\,e^{W(x)}=x$ is
known in mathematics as the Lambert $W$-function~\cite{Wfunction}. On
the negative real axis $W$ is real and two-valued over the interval
$-1/e\leq x \leq 0$, with $W(-1/e)=-1$, and  the two branches $W_0$ and
$W_{-1}$ are identified according to $W_0(x)\geq -1$ and $W_{-1}(x)\leq -1$. To
suppress $\langle D\rangle$ adequately we need to make $\alpha-W$
sufficiently large.  Given that $\frac{1}{c_D}\ll {\cal O}(1)$,
$\alpha=\frac{1}{c_D}+\frac{1}{2}$ is never large.  A large and
negative $W$ is in principle possible given that $W_{-1}(x)
\stackrel{x\to 0}{\longrightarrow}-\infty$. However, for small
negative values of its argument $\left|W_{-1}\right|$ grows only
logarithmically $W_1(-x)\approx \log x-\log(-\log x)$, and therefore an
exponential suppression of $\langle D\rangle$ would require an
exponentially small $\beta$.  Therefore, for two generations and
natural values of the parameters, no strong hierarchy can result.  We
can then conclude that in our scenario a scalar potential invariant under a
$SU(n_g)^5$~(flavor) symmetry can naturally yield a fully hierarchical
pattern of Yukawa couplings only when $n_g=3$.

\section{Spontaneous breaking of the Abelian symmetries}

What we have done until now can be easily generalized to include the
fields $N_{d,e}$ responsible for the breaking of the Abelian subgroups
$U(1)_{d,e}$.  Let us define for $i=d,\,e$ the $U(1)_i$ invariant
bilinears $T_{N_i}= N_iN_i^\dagger$.  These two invariants can couple
in a renormalizable way only among themselves or to the traces
$T_{u,d,e}$.  Omitting $A$, $D$, and logarithmic terms, by
generalizing the derivation of \eqn{eq:VTud} we can write the scalar
potential for the quadratic invariants ${\cal
  T}=(T_u,T_d,T_e,T_{N_d},T_{N_e})$ as
\begin{equation}
  \label{eq:VN}
  V_{\cal T} = 
\sum_{I J} \lambda_{IJ} 
\left[\left({\cal T}_I-\eta^2_I\right)+
\left({\cal T}_J-\eta^2_J\right)\right]^2\,, 
\end{equation}
where $\lambda_{IJ}= \lambda_{JI}$.  If all the eigenvalues of the
Hessian $\partial_I\partial_J V_{\cal T}$ are positive, the potential
is bounded and SSB occurs, resulting in the vevs $\langle {\cal
  T}_I\rangle =\eta^2_I$.  There is no need to give the explicit
relations between $\lambda_{IJ}$, $\eta^2_I$ and the parameters
$\lambda,\,m^2$ that appear in the potential when written in a more
familiar form, the important qualitative point is that one expects all
$\eta^2_I \sim {\cal O}(1)$. While on the one hand this means that the
inclusion of SSB of $U(1)_{d,e}$ leaves the results of the previous
sections unaffected, on the one other hand the ratios between the
bottom/tau masses and the mass of the top suggest instead $\eta_{N}
\sim 10^{-2}$.  It would be rather unpleasant, after reaching so far,
abandoning at this point our dogma about naturalness and assuming ad
hoc values for the fundamental parameters in order to reproduce these
two small numbers.  However, there are other possible ways out.  For
example, we can assign to the $SU(2)$ singlets $d$ and $e$ 
Abelian charges larger than one, which would imply a stronger
dimensional suppression of their effective Yukawa operators.  Taking for
example $\eta_{N}\sim 0.3 $, that we can still regard as a natural
value, then a charge of 4 would yield the required small factor
$\eta_{N}^4\sim 10^{-2}$.  Alternatively, we could simply assume that
the heavy messengers carrying $U(1)_{d,e}$ charges needed to generate
the effective Yukawa operators for $d$ and $e$ (see \eqn{eq:nonren})
have a mass scale $\Lambda'$ about two orders of magnitude larger than
the scale  $\Lambda$ of the neutral messengers. In this way $\eta_{N}
\approx (\Lambda/\Lambda')\; \eta_T \approx 10^{-2}$ is easily
obtained.

\section{Conclusions}

Let us recap what we have done and what we have achieved.  In the SM,
the fermions gauge invariant kinetic terms are characterized by a
$SU(5)^5$ quark/lepton flavor symmetry that is broken {\it explicitly}
by the Yukawa couplings. It is a big puzzle why the values of these
symmetry breaking parameters span six orders of magnitude, with no
apparent regularity besides the fact that their strong hierarchy is
qualitatively similar for all the three types of fermions.  We have
promoted this symmetry to an exact one, that is broken {\it
  spontaneously} by the ground state of a symmetry invariant scalar
potential.  We have chosen the multiplets of scalar fields in such a
way that their vevs correspond precisely to the Yukawa couplings so
that, among other things, by construction the SM Yukawa couplings are
the only source of flavor violation.  This promotes the spurion
technique (widely used in connection with the MFV hypothesis) to a
concrete piece of fundamental physics.  We have introduced a slight
variant with respect to the most popular MFV scenarios by factorizing
out the breaking of the two Abelian factors $U(1)_{d,e}$ from the
breaking of the non-Abelian $SU(5)^5$ flavor group.  This allowed us
to treat the lepton, up and down quark sectors in a similar way.  We
have first considered the three sectors uncoupled, and we have
parametrized the respective potentials in terms of the three TAD
invariants, which renders intuitively simple classifying the possible
structures of the symmetry breaking vacua.  We have found that for a
large part of the parameter space the ground state is characterized by
one component for each multiplet of Yukawa fields with an ${\cal
  O}(1)$ vev, while the vevs of the other components vanish.  To each
three-point and four-point scalar interactions we have then added a
symmetry invariant logarithmic correction, and we have shown that this
has the effect of lifting the vanishing vevs to exponentially
suppressed values. In this way a Yukawa hierarchy that is
qualitatively similar for all the three types of fermions arises quite
naturally.  As a further step we have included renormalizable terms
that couple the up and down quark sectors, and we have found that the
corresponding contributions to the scalar potential are minimized when
the heaviest (top and bottom) quarks are almost decoupled from the
lighter ones.  It is precisely the suppression of light-heavy mixings
that gives rise to the family structure of the quarks.  We have
briefly considered what would happen for a generic number of
generations, and we have concluded that for any number different from
three a fully hierarchical pattern cannot result.  This can be
restated in a stronger way by saying that, within our scenario, the
observed hierarchy of the Yukawa couplings implies precisely three
generations.  Finally, we have argued that accounting for the SSB of
the two $U(1)_{d,e}$ Abelian factors does not modify the previous
results. 

Clearly, several issues related with this work deserve further
studies, and some are listed below:

{\it Nambu-Goldstone bosons:}\ \ SSB implies the presence of
Nambu-Goldstone bosons (NGB), for which strong constrains exist from
unseen hadron decays, astrophysics, and from flavor violating
processes.  A standard solution is gauging the
symmetry~\cite{Albrecht:2010xh,Grinstein:2010ve}. A very large scale
suppressing the NGB couplings to ordinary particles could also provide
a way out.

{\it Leptons:}\ \ We have only considered the SM lepton sector, which
is known to be incomplete. It would be interesting to extend this
scenario to the most popular models for massive neutrinos, and
see if something could be said about neutrino masses and mixings.

{\it CKM matrix:}\ \ We have shown how the two heaviest quarks
naturally decouple from the lighter ones, hinting to the emergence of
a family structure for the quarks.  However, we have also concluded
that the full structure of $\langle V\rangle$ remains undetermined
because there is not enough information in the scalar potential we are
using. Clearly this point deserves further study.

{\it Effective potential:}\ \ What we consider by far the most
important issue is if the logarithmic terms that play such a crucial
role in our construction are effectively generated by quantum
corrections, and with reasonable values of the coefficients.  We
anticipate that, because of the several different types of
interactions and of the large number of component fields, this appears
to be a non-trivial task. However, once this program is carried out,
if the required logarithms will appear then we believe that a quite
interesting candidate for a theory of the SM Yukawa sector would have
been found.

% \vspace{-.3cm}

 \section*{Acknowledgments}

 It is a pleasure to thank Alberto Casas for encouraging comments and
 Jose Ramon Espinosa for pointing~out the complex nature of ${\cal
   D}$.  Rodrigo Alonso, Belen Gavela, Luca Merlo and Stefano Rigolin
 are warmly acknowledged for several conversations that stimulated
 this work.

%%%%%%%%%%%%%%%%%%%%%%%%%%%%%%%%%%%%%%%%%%%%%%%%%%%%%%%%%

%%%%%%%%%%%%%%%%%%%%%%%%%%%%%%%%%%%%%%%%%%%%%%%%%%%%%%%%%%%%%%%%%%%

\begin{thebibliography}{99}


\bibitem{MFV}
R.~S.~Chivukula, H.~Georgi,
  {\it Composite Technicolor Standard Model},
  Phys.\ Lett.\  {\bf B188}, 99 (1987). 
% L.~J.~Hall, L.~Randall,
%  {\it Weak scale effective supersymmetry},
%  Phys.\ Rev.\ Lett.\  {\bf 65}, 2939-2942 (1990).

% \bibitem{MFV2}
% A.~J.~Buras, P.~Gambino, M.~Gorbahn, S.~Jager and L.~Silvestrini,
% {\it Universal unitarity triangle and physics beyond the standard model},
%  Phys.\ Lett.\  B {\bf 500} (2001) 161
%  [arXiv:hep-ph/0007085].

%\cite{Joshipura:2009gi}
\bibitem{Joshipura:2009gi}
  A.~S.~Joshipura, K.~M.~Patel and S.~K.~Vempati,
  {\it Type I seesaw mechanism for quasi degenerate neutrinos}, 
  Phys.\ Lett.\  B {\bf 690}, 289 (2010)                        
  [arXiv:0911.5618 [hep-ph]].
  %%CITATION = PHLTA,B690,289;%%

%\cite{Alonso:2011jd}
\bibitem{Alonso:2011jd}
  R.~Alonso, G.~Isidori, L.~Merlo, L.~A.~Munoz, E.~Nardi,
  {\it Minimal flavour violation extensions of the seesaw},  
    [arXiv:1103.5461 [hep-ph]].

\bibitem{DAmbrosio:2002ex}
  G.~D'Ambrosio, G.~F.~Giudice, G.~Isidori and A.~Strumia,
  {\it Minimal flavour violation: An effective field theory approach},
  Nucl.\ Phys.\  B {\bf 645}, 155 (2002)
  [arXiv:hep-ph/0207036].
  %%CITATION = NUPHA,B645,155;%%

%\cite{Froggatt:1978nt}
\bibitem{Froggatt:1978nt}
  C.~D.~Froggatt, H.~B.~Nielsen,
  {\it Hierarchy of Quark Masses, Cabibbo Angles and CP Violation}, 
  Nucl.\ Phys.\  {\bf B147}, 277 (1979).

%\cite{Feldmann:2009dc}
\bibitem{Feldmann:2009dc}
  T.~Feldmann, M.~Jung, T.~Mannel,
  {\it Sequential Flavour Symmetry Breaking},
  Phys.\ Rev.\  {\bf D80}, 033003 (2009) 
  [arXiv:0906.1523 [hep-ph]].

%\cite{Alonso:2011yg}
\bibitem{Alonso:2011yg}
  R.~Alonso, M.~B.~Gavela, L.~Merlo, S.~Rigolin,
  {\it On The Potential of Minimal Flavour Violation}, 
    [arXiv:1103.2915 [hep-ph]].

%\cite{Nambu:1992ke}
\bibitem{Nambu:1992ke} Y.~Nambu, {\it Spontaneous symmetry breaking
    and the origin of mass}, Invited talk at the International
  Conference on fluid mechanics and theoretical physics in honor of
  Prof. Pei-Yuan Chou's 90th Anniversary, June 1-3, Beijing, China. 
Preprint EFI-92-37,
 %%CITATION = C92-06-01;%%

\bibitem{Wfunction}
R. M. Corless, G. H. Gonnet, D. E. G. Hare, D. J. Jeffrey and D. E. Knuth, 
{\it On the Lambert W Function}, 
Adv. Comp. Math., Vol. 5 (1996) 329.

%\cite{Albrecht:2010xh}
\bibitem{Albrecht:2010xh}
 M.~E.~Albrecht, T.~.Feldmann, T.~Mannel,
{\it Goldstone Bosons in Effective Theories with Spontaneously 
Broken Flavour Symmetry}, 
 JHEP {\bf 1010 } (2010)  089 [arXiv:1002.4798 [hep-ph]]; 
M.~E.~Albrecht,
{\it Two Approaches towards the Flavour Puzzle},
  PhD Thesis, Techn.\ Univ.\ Munich (2010), 
{\tt 
http://nbn-resolving.de/urn/resolver.pl?urn:nbn: de:bvb:91 -diss-20100818-981978-1-7}.

%\cite{Grinstein:2010ve}
\bibitem{Grinstein:2010ve} B.~Grinstein, M.~Redi and G.~Villadoro,
  {\it Low Scale Flavor Gauge Symmetries}, JHEP {\bf 1011}, 067 (2010)
  [arXiv:1009.2049 [hep-ph]].
  %%CITATION = JHEPA,1011,067;%%

\end{thebibliography}
   \end{document}